\documentclass[twocolumn,twocolappendix]{aastex63}

\received{}
\revised{}
\accepted{}
\submitjournal{ApJ}

\shorttitle{Negative Energy Surface Waves}
\shortauthors{Yu \& Nakariakov}
\begin{document}

\title{Excitation of negative energy surface magnetohydrodynamic waves in an incompressible cylindrical plasma}

\correspondingauthor{D. J. Yu}
\email{djyu79@gmail.com}

\author[0000-0003-1459-3057]{D. J. Yu}
\affiliation{Department of Astronomy and Space Science, Kyung Hee University,
 1732, Deogyeong-daero, Yongin, Gyeonggi 17104, South Korea}
\affiliation{School of Space Research, Kyung Hee University,
 1732, Deogyeong-daero, Yongin, Gyeonggi 17104, South Korea}

\author[0000-0001-6423-8286]{V. M. Nakariakov}
\affiliation{School of Space Research, Kyung Hee University,
 1732, Deogyeong-daero, Yongin, Gyeonggi 17104, South Korea}
 \affiliation{St. Petersburg Branch, Special Astrophysical Observatory, Russian Academy of Sciences, 196140, St. Petersburg, Russia}

\begin{abstract}
Negative energy wave phenomena may appear in shear flows in the presence of a wave decay mechanism and external energy supply. We study the appearance of negative energy surface waves in a plasma cylinder in the incompressible limit. The cylinder is surrounded by an axial magnetic field and by a plasma of different density. Considering flow inside and viscosity outside the flux tube, we derive dispersion relations, and obtain analytical solutions for the phase speed and growth rate (increment) of the waves. It is found that the critical speed shear for the occurrence of the dissipative instability associated with negative energy waves (NEWs) and the threshold of Kelvin--Helmholtz instability (KHI) depend on the axial wavelength. The critical shear for the appearance of sausage NEW is lowest for the longest axial wavelengths, while for kink waves the minimum value of the critical shear is reached for the axial wavelength comparable to the diameter of the cylinder. The range between the critical speed of the dissipative instability and the KHI threshold is shown to depend on the difference of the Alfv\'{e}n speeds inside and outside of the cylinder. For all axial {wavenumbers}, NEW appears for the shear flow speeds lower than the KHI threshold. It is easier to excite NEW in an underdense cylinder than in an overdense one. The negative energy surface waves can be effectively generated for azimuthal number $m=0$ with a large axial wave number and for higher modes ($m>0$) with a small axial wave number.

\end{abstract}

\keywords{magnetohydrodynamics(MHD) -- waves -- Sun: oscillations --  Sun: atmosphere}

\section{Introduction} \label{sec:intro}
The solar atmosphere is a highly structured and dynamic medium with pronounced non-uniformities of the macroscopic parameters of the plasma, such as the density and temperature, and the magnetic field, and also with a number of transient and long-living plasma flows. The non-uniform nature of the atmosphere strongly affects the propagation of magnetohydrodynamic (MHD) waves, leading to the appearance of the wave dispersion, enhanced damping, mode coupling, amplification and many other effects ~\citep[e.g.,][and references therein]{1983SoPh...88..179E, 2002ApJ...577..475R, 2012RSPTA.370.3193D,
2016GMS...216..449J, 2016GMS...216..431V, 2019FrASS...6...20G}. In particular, shear flows may greatly modify magnetohydrodynamic (MHD) wave propagation~\citep[e.g.,][]{1992SoPh..138..233G,1995SoPh..159..213N}.
Strong shear can induce Kelvin--Helmholtz instability (KHI)~\citep{1961hhs..book.....C,2015ApJ...813..123Z}. In addition, the effect of overreflection may occur for MHD waves reflected from a plasma non-uniformity with a velocity shear~\citep{1963PhFl....6..508F,1963PhFl....6..1154F,1970P&SS...18....1M, 1994AstL...20..763N, 2014JPlPh..80..667G}. In overreflection, the amplitude of the reflected wave is higher than the amplitude of the incident wave, i.e., the wave gains energy from the shear flow, or, more correctly, from the source that supports the shear flow. Overreflection is related to a backward nature of the transmitted wave, which changes the sign of the phase velocity due to the velocity shear, and its energy becomes negative. A wave with negative energy is called a negative energy wave (NEW) \citep[see, e.g.,][]{1979JFM....92....1C}. The use of this term emphasises that the amplitude of a NEW increases when the energy of the system decreases, for example, due to dissipative and wave leakage processes \citep[e.g.,][]{1986SvPhU..29.1040O, 1989SvPhU..32..783S}. It leads to the occurrence of various NEW instabilities. Nonlinear coupling of NEWs with regular waves with positive energy can become subject to explosive instabilities which are faster than the standard linear instabilities. MHD waves of negative energy attract attention in the context of the stability of shear flows in natural and laboratory plasmas \citep[e.g.,][]{2008PhPl...15e4501K, 2009NucFu..49c5008I}, including the solar atmosphere \citep[e.g.,][]{1997SoPh..176..285J, 2001A&A...368.1083A, 2011SSRv..158..505T, 2015A&A...577A..82B}.

An important feature of NEWs is that they can be unstable for the velocity shear well below the KHI threshold.  If the NEW instability is caused by a dissipative process, this phenomenon is called dissipative instability~\citep{1979JFM....92....1C,1997SoPh..176..285J}. \cite{1979JFM....92....1C} considered two parallel flows with {a} perpendicular profile of the velocity in {the} form of a step function and viscosity in one side, and proposed a criterion for the dissipative instability to occur at a shear flow speed below the KHI threshold. Despite the obvious importance of the effect of NEWs for the plasma non-uniformities of the solar atmosphere, there have been only several dedicated studies of this phenomenon.  Adopting~\cite{1979JFM....92....1C}'s criterion scheme,~\cite{1997SoPh..176..285J} studied the excitation condition for NEWs in a plasma slab, and showed that surface kink modes with negative energy could occur in magnetic structures of the solar photosphere. \cite{1988JETP...67.1594R}, considering kink modes in the long wavelength limit, was first to show that NEW may be crucial for the energy transfer to the upper solar atmosphere.
\cite{1995JPlPh..54..149R} obtained analytical solutions for the negative energy Alfv\'{e}n surface wave propagating on a discontinuous shear flow boundary in an incompressible plasma, taking into account viscosity at the one side and a constant flow at the other side. They showed that when the flow speed is above the critical value, one wave mode of two solutions changes the sign of the phase speed, and then two wave modes become co-propagating. The wave mode with the smaller phase speed has negative energy. Its growth rate (increment) increases with the increase in the viscosity coefficient. Recent study by \cite{2018JPlPh..84a9001R} showed that the growth rate of a standing surface wave is equal to the growth rate of the {(backward)} propagating wave with negative energy minus the damping rate (decrement) of the {(forward)} propagating wave with positive energy.

In this paper, we investigate the appearance of a negative energy MHD surface mode with an arbitrary azimuthal wave number $m$ in a plasma cylinder penetrated by an axial magnetic field, in the incompressible approximation.
We describe the model in Sec.~\ref{sec:model}, presents the results in Sec.~\ref{sec:results}, and conclude the paper in Sec.~\ref{sec:conclusion}

\section{Model} \label{sec:model}
Our governing equations are the viscous MHD equations for an incompressible plasma:
\begin{eqnarray}
\frac{\partial \rho}{\partial t}+\nabla\cdot(\rho\textbf{v})&=&0,\label{eq:1}\\
\rho\frac{\partial \textbf{v}}{\partial t}+\rho\textbf{v}\cdot\nabla\textbf{v}+\nabla p-\textbf{j}\times\textbf{B}&=&\eta\nabla^2\textbf{v},
\label{eq:2}\\
\frac{\partial p}{\partial t}+\textbf{v}\cdot\nabla p+\gamma p\nabla\cdot\textbf{v}&=&0,\label{eq:3}\\
\frac{\partial \textbf{B}}{\partial t}+\nabla\times\textbf{E}&=&0,\label{eq:4}\\
\textbf{j}-\frac{1}{\mu_0}\nabla\times\textbf{B}&=&0,\label{eq:5}\\
\textbf{E}+\textbf{v}\times\textbf{B}&=&0,\label{eq:6}
\end{eqnarray}
where $\textbf{v}$ is the velocity; $\textbf{B}$ and $\textbf{E}$ are the magnetic and electric fields, respectively; $\textbf{j}$ is the electric current density; $\rho$ is the mass density, $\eta$ is the shear viscosity, $\mu_0$ is the permeability of vacuum, and $\gamma$ is the ratio of specific heat.
We consider an infinitely long, axisymmetric cylindrical magnetic flux tube with radius $R$, i.e., a plasma cylinder with a sharp boundary, surrounded by a plasma with different physical quantities, similar to the model of \citet{1983SoPh...88..179E}. The magnetic field is parallel to the axis of the cylinder. The equilibrium is reached by the balance of the total pressure inside and outside the cylinder. Inside the cylinder there is a field aligned steady flow, uniform in the radial direction. The external plasma is static. Thus, the boundary of the cylinder is a tangential discontinuity. The plasma outside the cylinder has finite viscosity, while the internal plasma is ideal.

In the following, we linearized Eqs.~(\ref{eq:1})--(\ref{eq:6}) in cylindrical {coordinates}. We consider separately the regions inside and outside the cylinder as homogeneous media.  Perturbations are considered to be harmonic in time and with respect to the axial and azimuthal coordinates. Then, applying matching condition at the tube boundary, we derive {the} dispersion relation and obtain the solutions for the phase speed and damping or growth rate.

\subsection{Wave equations}\label{sec:wave-equation}
We denote the quantities inside (outside) of the cylinder by a subscript $i(e)$.
In the equilibrium, inside the cylinder, the magnetic field is $\textbf{B}_0=(0,0,B_{0})$, and the flow is $\textbf{v}_0=(0,0,U_{0})$. Both $U_0$ and $B_{0}$ are constants. The plasma density $\rho_i$ is constant too. The Alfv\'{e}n speed is $v_\mathrm{Ai}=B_{0}/\sqrt{\mu_0\rho_i}$. Linearizing {the} ideal MHD equations with respect to the equilibrium, and applying the Fourier transformation ($\sim\exp{[i(k_zz+m\phi-\omega t)]}$), we obtain for the perturbations of the radial velocity $\hat{v}_{ri}$ and total pressure $\hat{P}_{i}$ the following set of coupled ordinary differential equations (see Appendix \ref{sec:append_a}),
\begin{eqnarray}
-\rho_i(\tilde{\omega}^2
-\omega_\mathrm{Ai}^2)\hat{v}_{ri}
&=&i\tilde{\omega}\hat{P}_i',\label{eq:7}\\
i\bigg[\tilde{\omega}^2-v_\mathrm{Ai}^2\bigg(k_z^2+\frac{m^2}{r^2}\bigg)\bigg]
\tilde{\omega}\hat{P}_i&=&\rho_iv_\mathrm{Ai}^2(\tilde{\omega}^2-\omega_\mathrm{Ai}^2)
\frac{(r\hat{v}_{ri})'}{r},\nonumber\\\label{eq:8}
\end{eqnarray}
where $\tilde{\omega}=\omega-k_zU_0$, $\omega_\mathrm{Ai}=k_zv_\mathrm{Ai}$, and the prime denotes the derivative with respect to $r$. Here $k_z$ and $m$ are real, while $\omega$ could be complex.

Outside of the cylinder, we assume the same magnetic field as inside it, while the density is $\rho_e$.
Linearization of Eqs.~(\ref{eq:1})-(\ref{eq:6}) with the same Fourier transform leads to (see Appendix \ref{sec:append_b})
\begin{eqnarray}
\bigg({L}_z^2+\frac{4m^2\rho_e^2\nu_e^2\omega^2}{r^4}\bigg)\hat{v}_{re}
&=&\frac{2m^2\rho_e\nu_e\omega^2}{r^3}\hat{P}_e+i\omega{L}_z\hat{P}_e'\nonumber\\
&&-\frac{6im\rho_e^2\nu_e^2\omega^2}{r^4}\hat{v}_{\phi e},\label{eq:9}\\
i\omega\bigg({L}_z+\frac{m^2\rho_ev_\mathrm{Ae}^2}{r^2}\bigg)\hat{P}_e
&=&\rho_e v_\mathrm{Ae}^2\frac{(r\hat{v}_{re})'}{r}\nonumber\\
&&+\frac{2im^2\rho_e^2\nu_e\omega v_{Ae}^2}{r^3}\hat{v}_{re}
\label{eq:10}
\end{eqnarray}
where
\begin{eqnarray}
{L}_z&=&-\rho_e\bigg[\omega^2-\omega_\mathrm{Ae}^2-i\nu_e\omega\bigg(\mathcal{D}-\frac{1}{r^2} \bigg)\bigg],\label{eq:11}
\end{eqnarray}
$\nu_e=\eta/\rho_e$, $v_\mathrm{Ae}=B_{0}/\sqrt{\mu_0\rho_e}$, and $\mathcal{D}\psi=\frac{(r\psi')'}{r}-\big(\frac{m^2}{r^2}
+k_z^2\big)\psi$ for $\psi$. Notice that $\hat{v}_{\phi}$ denotes the  perturbation of the azimuthal velocity.

\subsection{Dispersion relation}
\label{sec:2.2}
{Taking divergence of Eq.~(\ref{eq:2}) yields the condition $\mathcal{D}\hat{P}_{i(e)}=0$  (see, e.g., \cite{1995JPlPh..54..149R}), which has Bessel functions as solutions.}

In this study we consider surface magnetohydrodynamic modes,
\begin{eqnarray}
\hat{P}_{i}&=&A_i I_m(k_zr),\label{eq:12}\\
\hat{P}_{e}&=&A_e K_m(k_zr),\label{eq:13}\\
\hat{v}_{ri}&=&-i\omega\hat{\zeta}_{ri}+ik_zU_0\hat{\zeta}_{ri}
=-i\tilde{\omega}\hat{\zeta}_{ri},\label{eq:14}\\
\hat{v}_{re}&=&-i\omega\hat{\zeta}_{re},\label{eq:15}
\end{eqnarray}
where $A_i$ and $A_e$ are constant, $I_m(k_zr)$ and $K_m(k_zr)$ are modified Bessel functions of the first and second kinds, respectively, and $\hat{\zeta}_r$ is the Fourier-transformed Lagrangian displacement in the radial direction. Hereafter we use the notations $I_m$ and $K_m$ instead of $I_m(k_zr)$ and $K_m(k_zr)$, respectively.

From the kinematic boundary condition, and continuity condition of the stress tensor at the boundary ($r=R$),
we obtain
\begin{eqnarray}
\frac{\partial \hat{v}_z}{\partial r}&+&ik_z \hat{v}_r=0,\label{eq:16}\\
\hat{P}_i&=&\hat{P}_e-2\rho_e\nu_e\frac{\partial \hat{v}_{re}}{\partial r},\label{eq:17}\\
\hat{\zeta}_{ri}&=&\hat{\zeta}_{re}.\label{eq:18}
\end{eqnarray}

Next, substituting {Eqs. (\ref{eq:12})--(\ref{eq:17})} into Eqs.~(\ref{eq:7}) and (\ref{eq:9}), we derive
\begin{eqnarray}
\rho_i(\tilde{\omega}^2-\omega_\mathrm{Ai}^2)\hat{\zeta}_{ri}
&=&k_zA_iI_m',\label{eq:19}\\
-i\omega{L}_z^2\hat{\zeta}_{re}
&=&\frac{2m^2\rho_e\nu_e\omega^2}{r^3}A_eK_m +i\omega k_zA_e{L}_zK_m',\nonumber\\\label{eq:20}
\end{eqnarray}
where the prime denotes the derivative with respect to entire argument $k_zR$. In Eq.~(\ref{eq:20}), we neglect the term with $\nu_e^2$, assuming the viscosity to be weak, ${\nu_e/(\omega R^2)\ll1}$. This condition can be written as $R_ek_zR\gg1$, where $R_e=\omega R/k_z\nu_e$ is the Reynolds number. For sufficiently large values of $R_e$, this approximation is valid regardless of the value of $k_zR$ ($R_e\gg k_zR$). {In other words, this condition implies that there exists a lower limit (cutoff) to $k_zR$ for a given $\nu_e$ such that $k_zR\gg(1/R_e)$, which must be considered for the interpretation of the results in Figs.~(\ref{fig:fig7}) and (\ref{fig:fig8}) in Sec. \ref{sec:3.3}.}

Then we obtain for $A_i$ and $A_e$
\begin{eqnarray}
A_i&=&\frac{\rho_i}{k_z I_m'}(\tilde{\omega}^2-\omega_\mathrm{Ai}^2)\hat{\zeta}_{ri}
,\label{eq:21}\\
A_e&=&\frac{\rho_e\big[(\omega^2-\omega_\mathrm{Ae}^2)
-2i\nu_e\omega\big(\mathcal{D}-\frac{1}{R^2}\big)\big]\hat{\zeta}_{re}}
{k_zK_m'}\nonumber\\&&\times\bigg[1-\frac{i\nu_e\omega\big[\frac{2m^2}{R^3}K_m
-k_z\big(\mathcal{D}-\frac{1}{R^2}\big)K_m'\big]}{k_z(\omega^2-\omega_\mathrm{Ae}^2)K_m'}\bigg]
.~~\label{eq:22}
\end{eqnarray}
Combining Eqs.~(\ref{eq:15}) and (\ref{eq:17}) yields
\begin{eqnarray}
\hat{P}_i&=&\hat{P}_e+2i\omega\rho_e\nu_e\frac{\partial \hat{\zeta}_{re}}{\partial r}.\label{eq:23}
\end{eqnarray}
Substituting Eqs. (\ref{eq:12})-(\ref{eq:13}) with (\ref{eq:21})-(\ref{eq:22}) into Eq. (\ref{eq:23}), we obtain
\begin{widetext}
\begin{eqnarray}
&&\bigg[\bigg(\frac{\rho_iI_m}{k_z I_m'}-\frac{\rho_eK_m}{k_z K_m'}\bigg)\omega^2-2k_zU_0\frac{\rho_iI_m}{k_z I_m'}\omega+\frac{\rho_iI_m}{k_z I_m'}(k_zU_0)^2-\frac{\rho_iI_m}{k_z I_m'}\omega_\mathrm{Ai}^2+\frac{\rho_eK_m}{k_z K_m'}\omega_\mathrm{Ae}^2\bigg]\hat{\zeta}_{ri}\nonumber\\
&+&i\omega\rho_e\nu_e\bigg\{\frac{2K_m}{k_z K_m'}\bigg(\mathcal{D}-\frac{1}{r^2} \bigg)-\frac{K_m}{k_z K_m'^2}\bigg[\bigg(\mathcal{D}-\frac{1}{r^2} \bigg)K_m'\bigg]+\frac{2m^2K_m^2}{k_z^2R^3K_m'^2}-2\frac{\partial }{\partial r}\bigg\}\hat{\zeta}_{re}=0.\label{eq:24}
\end{eqnarray}
\end{widetext}
As we are interested in the behavior of the wave amplitude in time, we may assume that approximately
\begin{eqnarray}
\hat{\zeta}_{re}&\approx&\frac{1}{\rho_e(\omega^2-\omega_{Ae}^2)}\frac{\partial \hat{P}_{re}}{\partial r} =\frac{1}{\rho_e(\omega^2-\omega_{Ae}^2)}\frac{\partial (A_eK_m)}{\partial r}\nonumber\\
&=&\frac{k_zA_eK_m'}{\rho_e(\omega^2-\omega_{Ae}^2)}. \label{eq:25}
\end{eqnarray}
Then we obtain the dispersion relation, by using boundary condition~(\ref{eq:18}) {and Eq.~(\ref{eq:25})},
\begin{widetext}
\begin{eqnarray}
\bigg[\bigg(\frac{\rho_iI_m}{k_z I_m'}-\frac{\rho_eK_m}{k_z K_m'}\bigg)\omega^2&&-2k_zU_0\frac{\rho_iI_m}{k_z I_m'}\omega+\frac{\rho_iI_m}{k_z I_m'}(k_zU_0)^2-\frac{\rho_iI_m}{k_z I_m'}\omega_\mathrm{Ai}^2+\frac{\rho_eK_m}{k_z K_m'}\omega_\mathrm{Ae}^2\bigg]\nonumber\\
&&+i\omega\rho_e\nu_e\bigg[\frac{K_m}{k_z K_m'^2}\bigg(\mathcal{D}K_m'-\frac{K_m'}{R^2} \bigg)+\frac{2m^2K_m^2}{k_z^2R^3K_m'^2}-2k_z\frac{K_m''}{K_m'}\bigg]=0.\label{eq:26}
\end{eqnarray}
\end{widetext}
Equation (\ref{eq:26}) can be rewritten as
\begin{eqnarray}
{D}=a\omega^2&+&b\omega+c+id\omega=0,\label{eq:27}
\end{eqnarray}
which is a quadratic equation for $\omega$ with the coefficients
\begin{eqnarray}
a&=&\frac{I_m}{I_m'}-\frac{R_\mathrm{ei}K_m}{K_m'},\label{eq:28}\\
 b&=&-2k_zU_0\frac{I_m}{I_m'},\label{eq:29} \\
 c&=&k_z^2v_\mathrm{Ai}^2\frac{I_m}{I_m'}\bigg[\tilde{U}_0^2-\bigg(1-\frac{I_m'K_m}{ I_mK_m'}R_\mathrm{ei}V_\mathrm{ei}^2\bigg)\bigg], \label{eq:30}\\
 d&=&\nu_eR_\mathrm{ei}\bigg[\frac{K_m}{ K_m'^2}\bigg(\mathcal{D}K_m'-\frac{K_m'}{R^2} \bigg)-2k_z^2\frac{K_m''}{K_m'}+\frac{2m^2K_m^2}{k_zR^3K_m'^2}\bigg],\nonumber\\\label{eq:31}
\end{eqnarray}
and introducing dimensionless ratios $R_\mathrm{ei}=\rho_e/\rho_i$, ${\tilde{U}_0=U_0/v_\mathrm{Ai}}$, and $V_\mathrm{ei}=v_\mathrm{Ae}/v_\mathrm{Ai}$. When  $U_0=\eta=0$, Eq.~(\ref{eq:27}) reduces to the dispersion relation for MHD surface waves in the incompressible limit~\citep{1983SoPh...88..179E}.

Taking that $\omega=\omega_r+i\omega_i$, where $\omega_r$ and $\omega_i$ are the real and imaginary parts of the cyclic frequency, and assuming $\omega_i\ll\omega_r$, we reduce Eq.~(\ref{eq:27}) to

\begin{eqnarray}
(a\omega_r^2+b\omega_r+c)+i(2a\omega_r\omega_i+b\omega_i+d\omega_r)=0.\label{eq:32}
\end{eqnarray}
The general analytical solutions for $\omega_r$ and $\omega_i$ are then
\begin{eqnarray}
\omega_{r\pm}&=&\frac{-b\pm\sqrt{b^2-4ac}}{2a},\label{eq:33}\\
\omega_{i\pm}&=&\frac{-d\omega_r}{2a\omega_r+b}=\mp\frac{d\omega_r}{\sqrt{b^2-4ac}}.\label{eq:34}
\end{eqnarray}

The normalized phase speed $\tilde{v}_p$ and damping (or growth) rate $\gamma$ are consequently given as
\begin{eqnarray}
\tilde{v}_{p\pm}&=&\frac{\omega_{r\pm}}{k_zv_\mathrm{Ai}}\nonumber\\
&=&\frac{\tilde{U}_0\pm
\sqrt{\frac{R_\mathrm{ei}I_m'K_m}{ I_mK_m'}\tilde{U}_0^2+AB}}
{B},\label{eq:35}
\end{eqnarray}
and
\begin{equation}
\gamma_{\pm}=\frac{\omega_i}{\omega_r}\label{eq:36}=
\end{equation}
\begin{displaymath}
\pm\frac{\tilde{\nu}_eR_\mathrm{ei}\frac{I_m'}{I_m}\bigg[\frac{K_m\big(R^2\nabla^2K_m'-{K_m'} \big)}{ \tilde{k}_zK_m'^2}-2\tilde{k}_z\frac{K_m''}{K_m'}+\frac{2m^ 2K_m^2}{\tilde{k}_z^2K_m'^2}\bigg]}
{\sqrt{\frac{\rho_{ei}I_m'K_m}{I_m K_m'}\tilde{U}_0^2+AB}},\nonumber
\end{displaymath}
where $\tilde{\nu}_e=\nu_e/v_\mathrm{Ai}R$, $\tilde{k}_z=k_zR$, and
\begin{eqnarray}
A=\bigg(1-\frac{I_m'K_m}{I_mK_m'}R_\mathrm{ei}V_\mathrm{ei}^2\bigg),~~\nonumber
B=\bigg(1-\frac{R_\mathrm{ei}I_m'K_m}{ I_mK_m'}\bigg).\nonumber
\end{eqnarray}

The solutions of $c=0$ and ${b^2-4ac=0}$ correspond to the critical speed, $U_\mathrm{c}$, and the threshold for KHI, $U_\mathrm{KH}$, respectively~\citep[e.g.,][]{1995JPlPh..54..149R}:
\begin{eqnarray}
U_\mathrm{c}&=&v_\mathrm{Ai}\sqrt{1-\frac{I_m'K_m}{ I_mK_m'}R_\mathrm{ei}V_\mathrm{ei}^2},\label{eq:37}\\
U_\mathrm{KH}&=&\Big[-\frac{I_m K_m'}{R_\mathrm{ei}I_m'K_m}
\bigg(1-\frac{R_\mathrm{ei}I_m'K_m}{I_m K_m'}\bigg)\Big]^{1/2} U_\mathrm{c}.\label{eq:38}
\end{eqnarray}

We now establish the criterion for the negative energy wave excitation. Using the dispersion relation without viscosity, ${D}_0=a\omega^2+b\omega+c$, we can determine the criterion for the negative wave energy to exist as $C=\omega\frac{\partial D_0}{\partial\omega}<0$~\citep{1979JFM....92....1C,1997SoPh..176..285J}. In the absence of the steady flow, we need to have $C>0$ and find that $C$ is positive except $\omega=0$. In the presence of flow, from the condition $C<0$, we obtain the criterion for the occurrence of a negative energy MHD surface wave
\begin{eqnarray}
  0<\omega<\omega_N,~~\omega_N=-\frac{b}{2a}=
  \frac{k_zU_0}{1-\frac{R_\mathrm{ei}I_m'K_m}{I_mK_m'}}.\label{eq:39}
\end{eqnarray}
The condition $\omega(=\omega_r)>0$ yields the relation $U_0>U_\mathrm{c}$, which is consistent with previous results~\citep[e.g.,][]{1995JPlPh..54..149R}.

{We point out that if we ignore the term with the factor $d$ in Eq.~(\ref{eq:32}) (no viscosity), Eq.~(\ref{eq:33}) describes the KHI when $b^2-4ac<0$.}

\section{Results} \label{sec:results}

\subsection{Dispersion curves}
\begin{figure}
\includegraphics[width=0.45\textwidth]{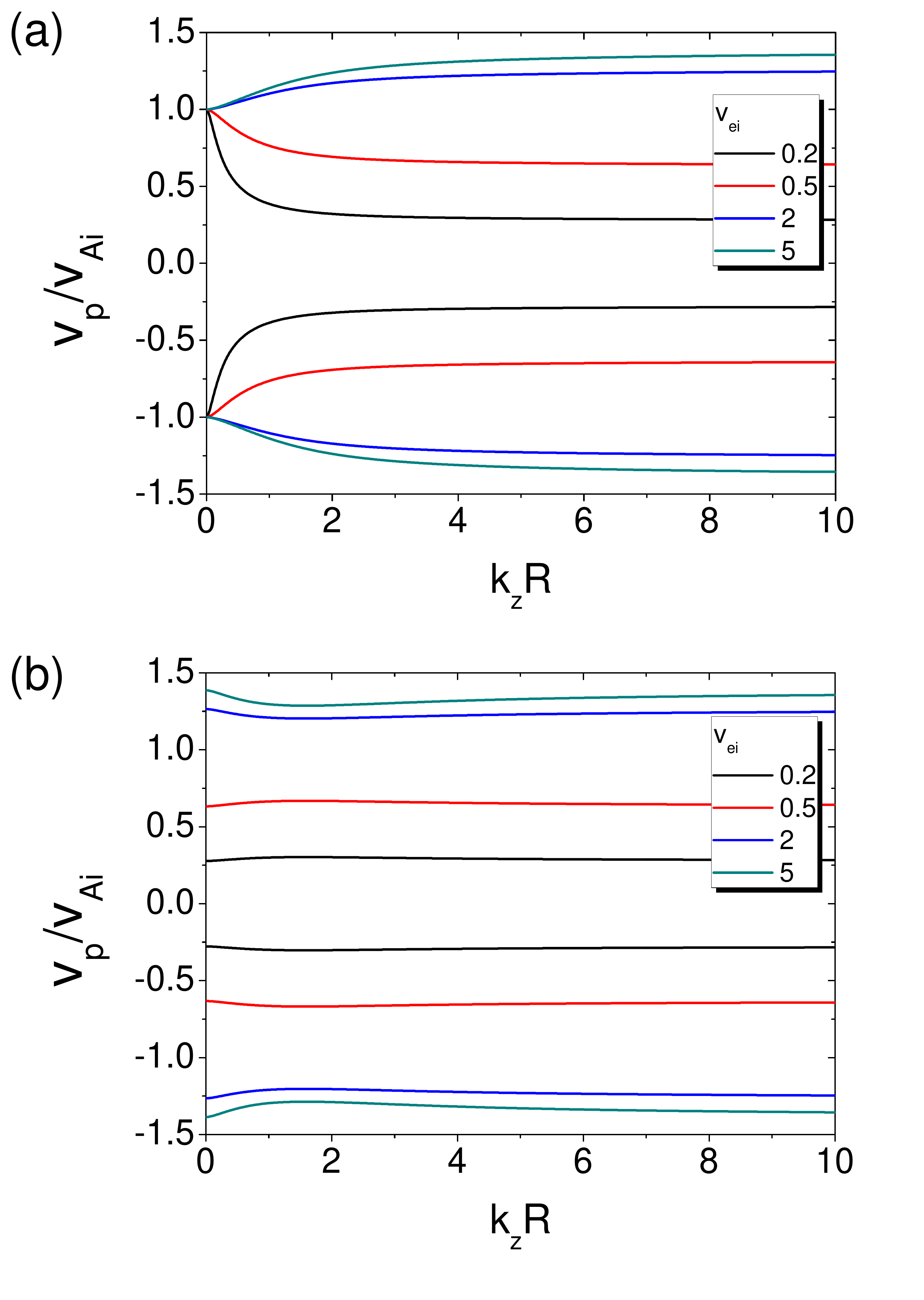}
\caption{\label{fig:fig1} Phase speeds of Alfv\'en surface modes of an incompressible plasma cylinder as a function of the axial wave number $k_zR$ for (a) $m=0$ and for (b) $m=1$ in the case with no steady flow $\tilde{U}_0=0$. }
\end{figure}

We first compare the dispersion curves for the phase speed with and without the steady flow. In Fig.~\ref{fig:fig1} we present the phase speed $v_\mathrm{p}$ as a function of $k_zR$ for two lowest azimuthal mode numbers and different values of the Alfv\'en speed ratios $V_\mathrm{ei}$ in a cylinder with $\tilde{U}_0=0$. The phase speed curves obtained in the cases $V_\mathrm{ei}>1$ and $V_\mathrm{ei}<1$ show different behaviour, which is consistent with the result obtained by \citet{1983SoPh...88..179E}. In the following, we shall denote the higher and lower phase speeds  as $v_{p+}$ and $v_{p-}$, respectively. In the static ($\tilde{U}_0=0$) case, the modes with $v_{p+}$ and $v_{p-}$ propagate in the positive and negative $z$-directions, and $v_{p+} = - v_{p-}$.  The dispersion curves for $m>1$ are similar to $m=1$ case, but as $m$ increases the curvature of the curves becomes flatter, i.e., the wave dispersion decreases.

\begin{figure}
\includegraphics[width=0.45\textwidth]{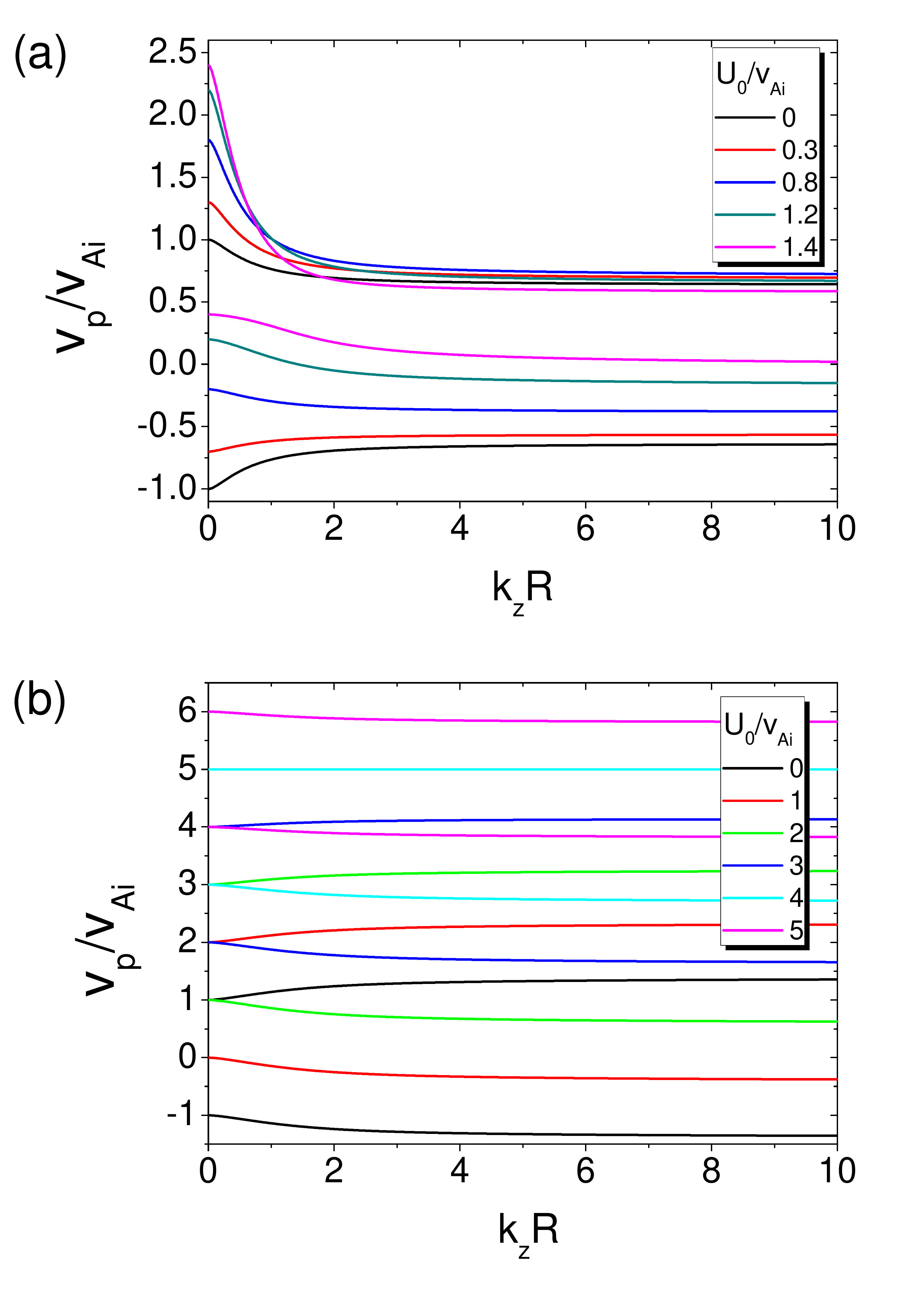}
\caption{\label{fig:fig2} Phase speeds of  the sausage mode ($m=0$) as a function of the axial wave number $k_zR$ for (a)  $V_\mathrm{ei}=0.5$, and (b) $V_\mathrm{ei}=5.0$ for different values of the steady flow speed $U_0$. }
\end{figure}

\begin{figure}
\includegraphics[width=0.45\textwidth]{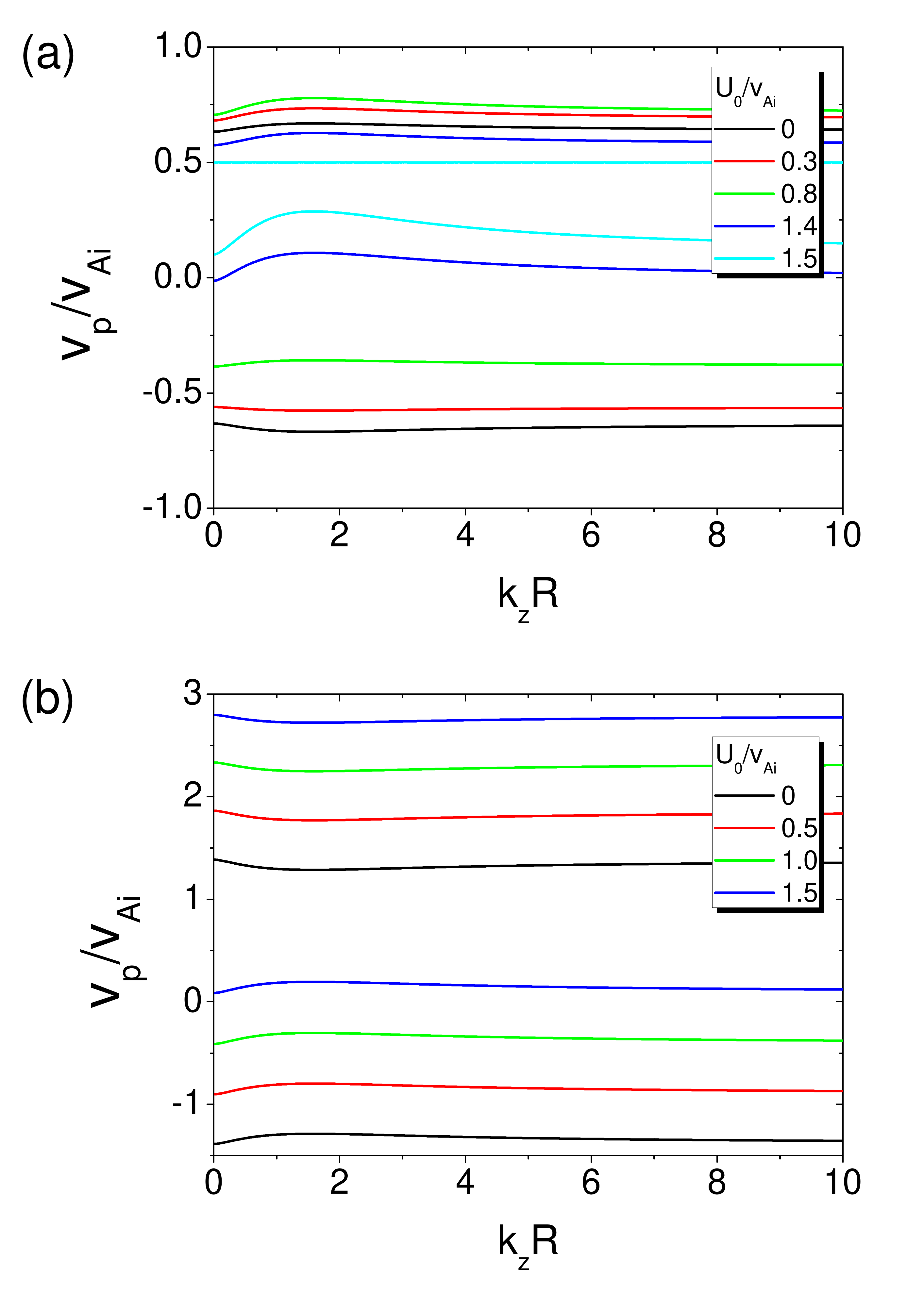}
\caption{\label{fig:fig3} The same as in Fig.~\ref{fig:fig2} but for the kink mode ($m=1$). }
\end{figure}

Fig.~\ref{fig:fig2} shows the dependence of dispersion curves for $m=0$ sausage modes on the steady flow speed for two different values of the Alfv\'en speed ratio $V_\mathrm{ei}$. In the presence of the steady flow $U_0$, the symmetry of the waves propagating in the opposite directions is broken, i.e., the values of $v_{p+}$ and $v_{p-}$ are affected by the steady flow differently. It is consistent with the result obtained by \citet{1997SoPh..176..285J}. For a sufficiently large $U_0$ both $v_{p+}$ and $v_{p-}$ are positive. In this regime, the $v_{p-}$ mode {becomes} a backward wave.
For $V_\mathrm{ei} < 1$, as the steady flow speed $U_0$ increases, the curve for $v_{p+}$ shifts first upward and then shifts downward, which {does not happen} for $v_{p-}$. The curve for $v_{p-}$ goes upward as $U_0$ increases. On the other hand, for $V_\mathrm{ei}=5$, both curves go up with $U_0$ increment. The same behavior is found for the $m=1$ kink modes (see Fig.~\ref{fig:fig3}). The feature of backward shift for $v_{p+}$ appears to be common for $V_\mathrm{ei}<1$, for both $m=0$ and $m=1$ modes.

\subsection{Conditions for the excitation of negative energy waves}

\begin{figure}
\includegraphics[width=0.45\textwidth]{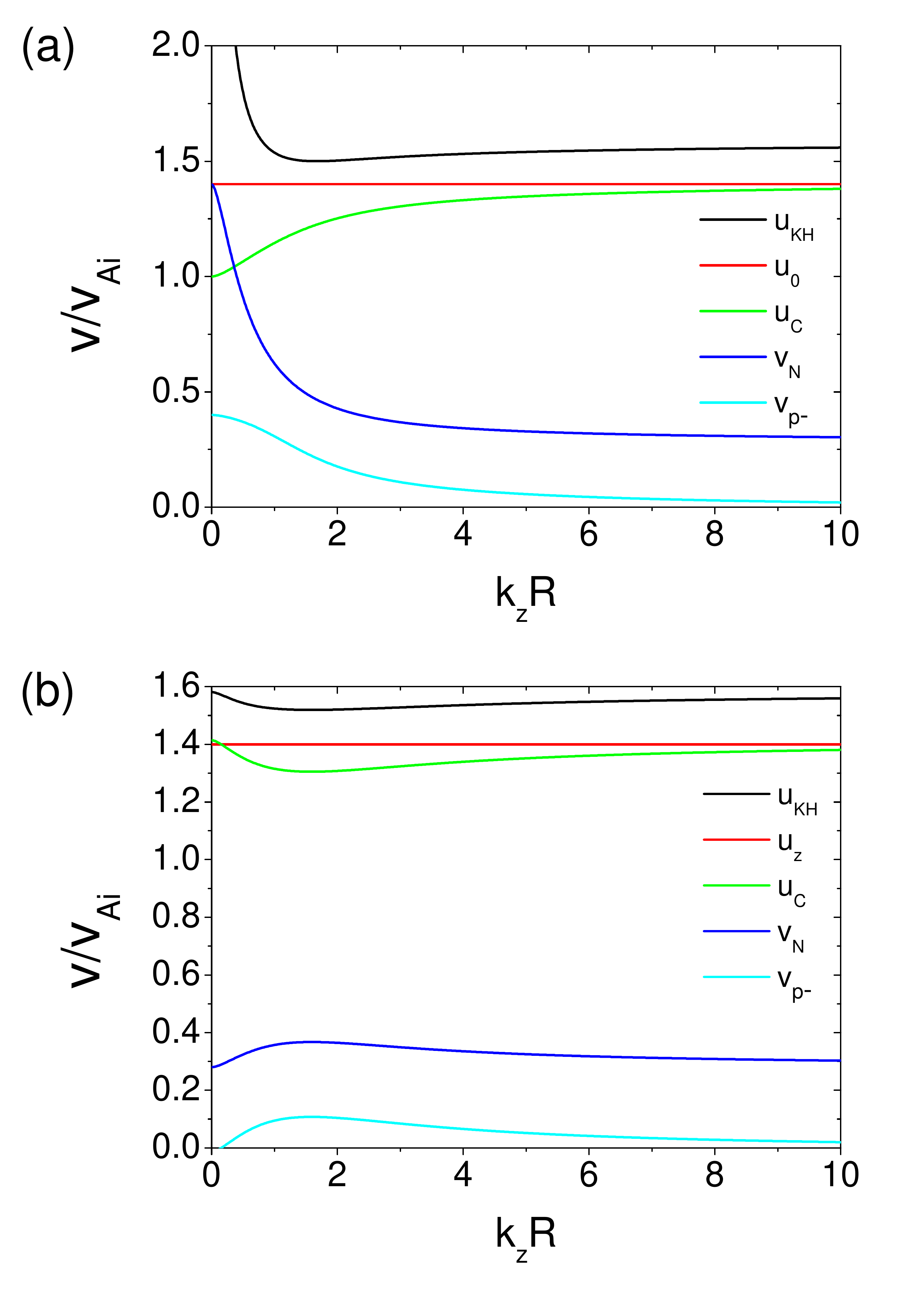}
\caption{\label{fig:fig4} Dependence of the characteristic speeds $v_\mathrm{p-}$, $v_\mathrm{N}$, $U_0$, $U_\mathrm{KH}$, and $U_\mathrm{c}$ on the axial wave number $k_zR$ for (a) $m=0$ and (b) $m=1$ modes, for $U_0/v_\mathrm{Ai}=1.4$ and $V_\mathrm{ei}=0.5$. }
\end{figure}

From Eq.~(\ref{eq:39}) we infer that in the backward regime, when $v_\mathrm{p-}>0$, a mode with $v_\mathrm{p-}$ can become a NEW. In Fig.~\ref{fig:fig4} we plot several characteristic speeds  ($v_\mathrm{p-}$, $v_\mathrm{N}(=\omega_\mathrm{N}/k_zv_\mathrm{Ai})$, $U_\mathrm{c}$, and $U_\mathrm{KH}$), as a function of $k_zR$ for (a) $m=0$ and (b) $m=1$, for fixed $\tilde{u}_z(=U_0/v_\mathrm{Ai})$ and $V_\mathrm{ei} < 1$. As inspected in the previous section, it is shown that the condition $U_0>U_\mathrm{c}$ corresponds to $v_\mathrm{p-} > 0$ and $v_\mathrm{p-} < v_\mathrm{N}$. When $\tilde{U}_0$ is over 1 for the sausage mode and 1.3 for the kink mode, the backward wave become a NEW, and hence its amplitude can grow exponentially due to one of the NEW instabilities (see Fig.~\ref{fig:fig7} (c)). When $U_0$ is sufficiently large, the KHI threshold could be reached for certain values of the axial wave number. For {the} $m=0$ mode, $U_\mathrm{KH}$ approaches infinity as $k_zR$ goes to zero, so it is not possible for the sausage surface mode to be KH unstable in the long wavelength limit. Other modes have finite values of $U_\mathrm{KH}$.

\begin{figure}
\includegraphics[width=0.45\textwidth]{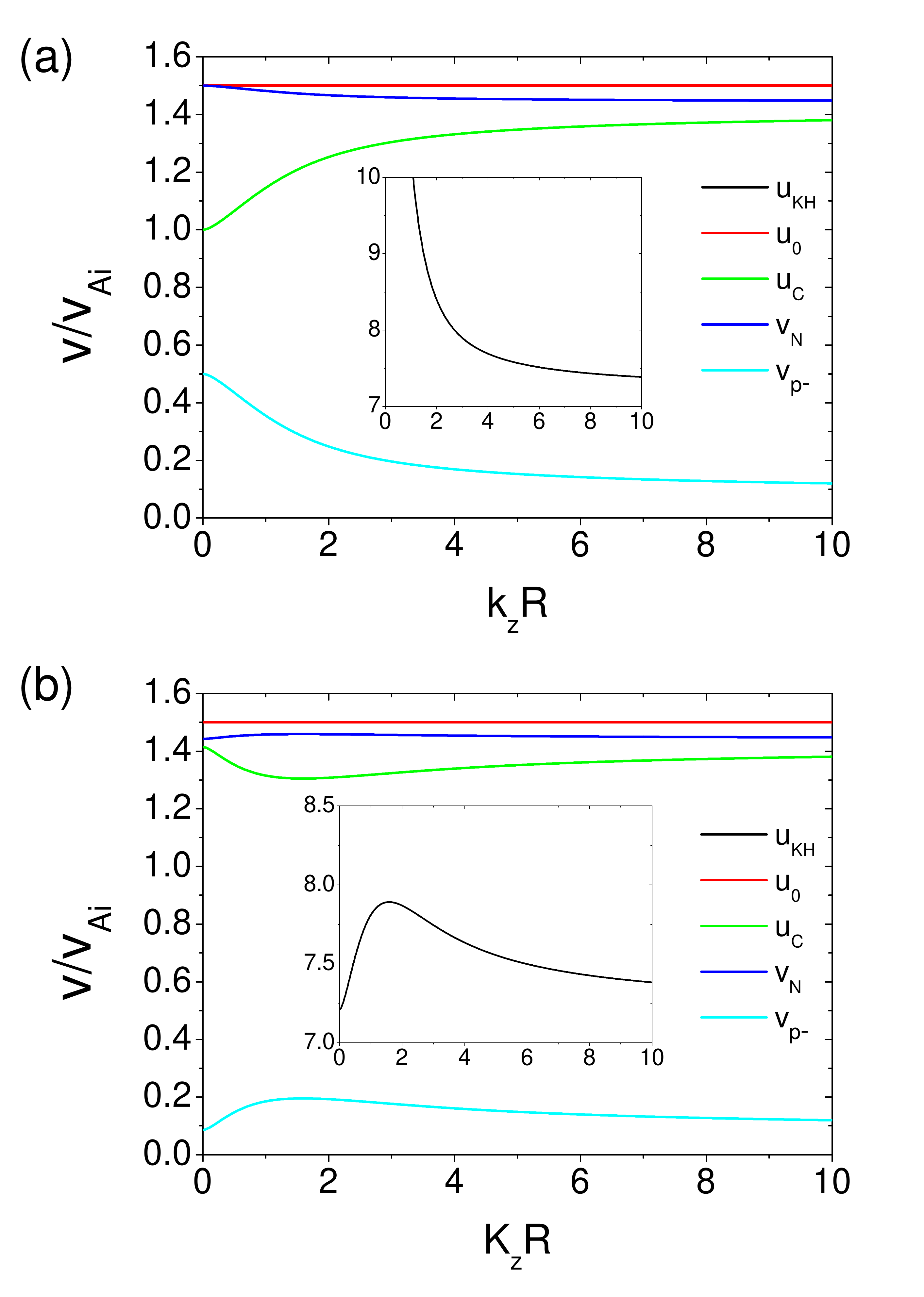}
\caption{\label{fig:fig5} The same as in Fig.~\ref{fig:fig4}, but for $U_0/v_\mathrm{Ai}=1.5$ and $V_\mathrm{ei}=5$ }
\end{figure}

For $V_\mathrm{ei} > 1$, the picture slightly changes, see Fig~\ref{fig:fig5}. The value of $U_\mathrm{KH}$ becomes large depending on $R_\mathrm{ei}(V_\mathrm{ei})$ (see Eq.~(\ref{eq:38})). The shape of the dependence of  $v_\mathrm{N}$ also greatly changes. The shape of the $U_\mathrm{c}$ curve does not change significantly.

\begin{figure}
\includegraphics[width=0.45\textwidth]{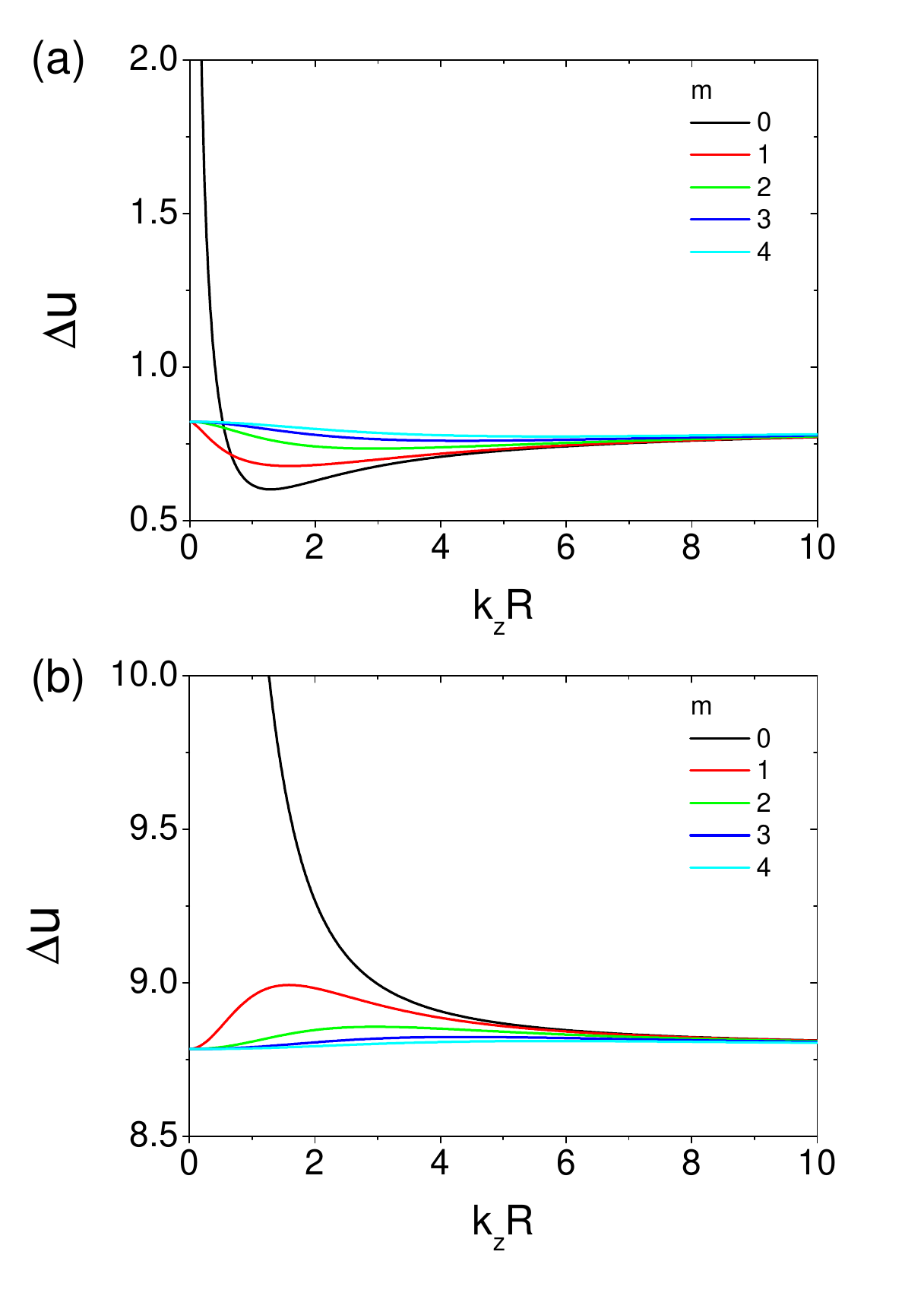}
\caption{\label{fig:fig6} The difference between the thresholds of KHI and NEW instabilities $\Delta U(=(U_\mathrm{KH}-U_\mathrm{c})/v_\mathrm{Ai})$ vs the axial wave number $k_zR$ for (a) $V_\mathrm{ei}=0.5$ and (b) $V_\mathrm{ei}=5$. }
\end{figure}

As $U_\mathrm{KH}$ and $U_\mathrm{c}$ both changes with $V_\mathrm{ei}$, the interval of the values of $U_0$ in which NEW instabilities are possible needs to be specified. In Fig.~\ref{fig:fig6}, we plot $\Delta U \equiv (U_\mathrm{KH}-U_\mathrm{c})/v_\mathrm{Ai})$ versus $k_zR$ for (a) $V_\mathrm{ei} < 1$ and (b) $V_\mathrm{ei} > 1$. For positive $\Delta U$, NEW instabilities have lower threshold than KHI.  The dependences of $\Delta U$ on the axial wave number are different in the cases of  $V_\mathrm{ei}<1$ and $V_\mathrm{ei}>1$. An increase in the Alfv\'en speed ratio $V_\mathrm{ei}$ increases the KHI threshold, implying NEW instabilities are more likely.

\subsection{Growth rate for the negative energy MHD modes}
\label{sec:3.3}
\begin{figure}
\includegraphics[width=0.45\textwidth]{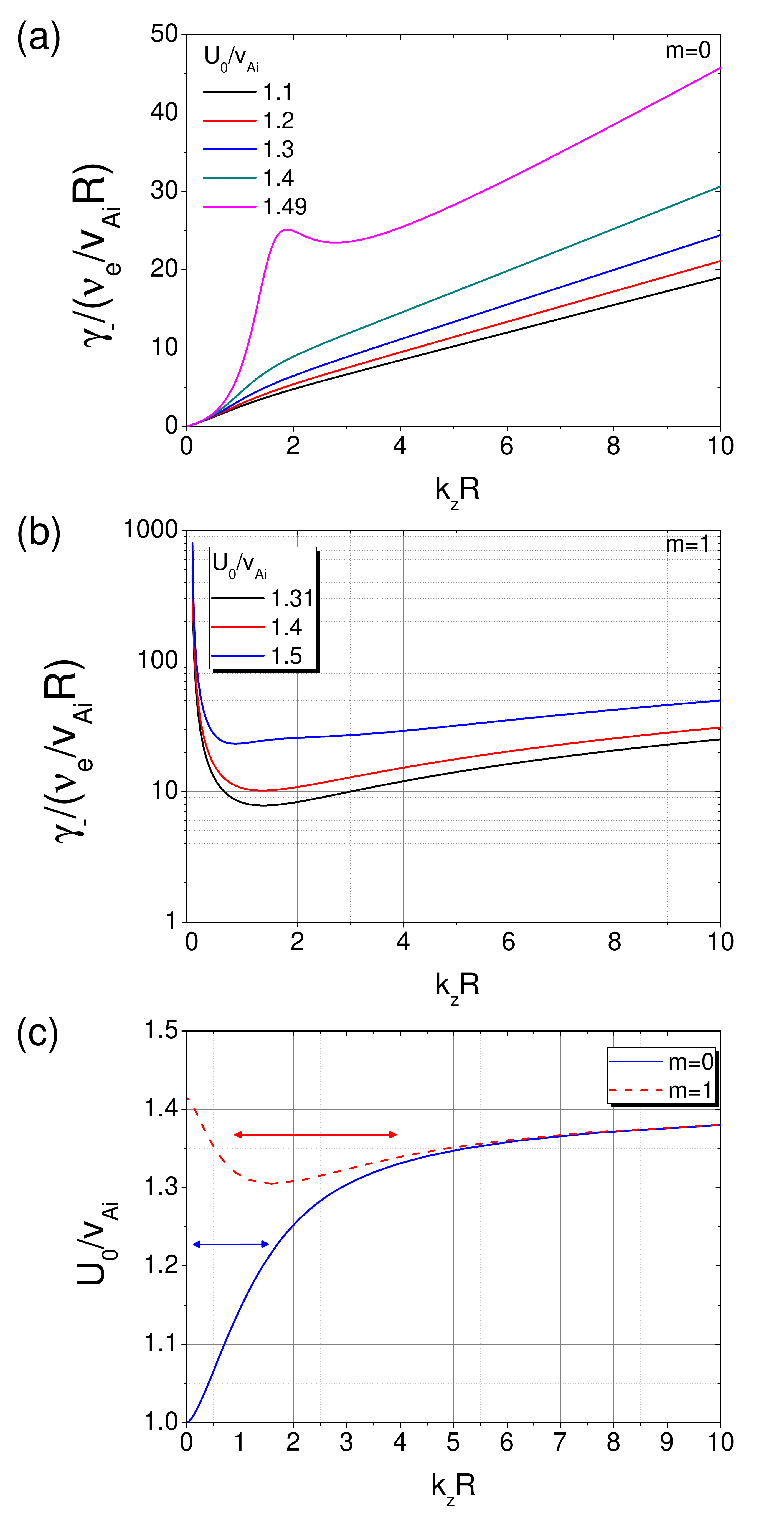}
\caption{\label{fig:fig7} The curve of $\gamma_{-}/\tilde{\nu}_e$ vs $k_zR$ for (a) $m=0$ and (b) $m=1$ modes, and (c) the range of $k_zR$ (denoted by the arrows) for the existence of NEW instabilities for  $V_\mathrm{ei}=0.5$ and $\tilde{\nu}_e=0.0001$. }
\end{figure}

In the NEW regime, the finite viscosity leads to the amplification of the waves, which is characterised by the imaginary part of the frequency, given by Eq.~(\ref{eq:36}).  {In Fig.~\ref{fig:fig7} (a), (b), we plot $\gamma_{-}/\tilde{\nu}_e$, the growth rate $\gamma_{-}$ of the backward wave divided by $\tilde{\nu}_e$,} versus $k_zR$ for the sausage and kink modes for different shear flow speeds $\tilde{U}_0$ for both $V_\mathrm{ei} < 1$ and $\tilde{\nu}=0.0001$. As $\tilde{U}_0$ increases, the values of $\gamma_-$ grows, and its value is higher for the kink mode than for the sausage mode. The growth rate of the kink mode has a minimum near $k_zR=1$, with its specific value depending on $U_0$.  In contrast, the growth rate of the sausage mode may have a local maximum, while {it} keeps growing with the increase in $k_zR$.

In Fig.~\ref{fig:fig7} (c) we show the range of $k_zR$ for the existence of NEW instabilities, which is obtained by applying the condition $U_0=U_\mathrm{c}$. For the sausage mode, the range starts from $k_zR=0$ where $\tilde{U}_0=1$ and extends to larger values of $k_zR$ as $\tilde{U}_0$ increases. On the other hand, for the kink mode, the NEW unstable range starts from the point $k_zR\approx1.58$ for the used plasma parameters, and becomes wider with the increase in $U_0$. For $m>0$, it is found that as $m$ increases, $\gamma_-$ increases in the whole range of $k_zR$. From the results, one may anticipate that the sausage mode is most unstable to NEW instability when $k_zR$ is sufficiently large while higher modes are most unstable in $k_zR\approx0$. The growth rate for high-$m$ modes becomes large when $k_zR$ approaches zero, which violates the assumption $\omega_i\ll\omega_r$.

\begin{figure}
\includegraphics[width=0.45\textwidth]{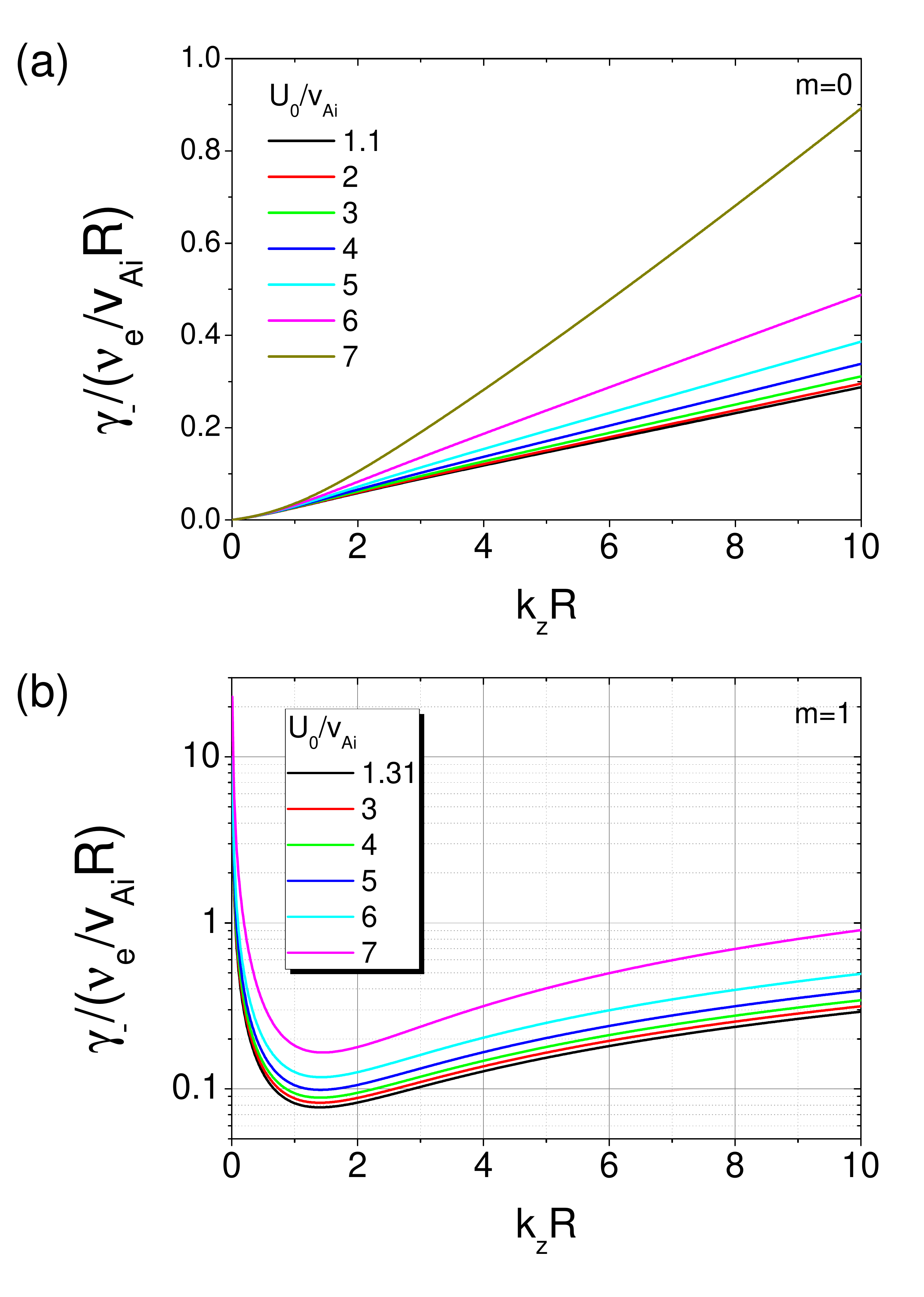}
\caption{\label{fig:fig8} The same as in Fig.~\ref{fig:fig7} (a) and (b), but for $V_\mathrm{ei}=5$. The range of $k_zR$ for the excitation condition of NEW instabilities is the same as Fig.~\ref{fig:fig7} (c). }
\end{figure}

The behavior of $\gamma_-$ for $V_\mathrm{ei} > 1$ is presented in Fig.~\ref{fig:fig8}. The growth rate is much lower than in Fig.~\ref{fig:fig7}, implying that an overdense flux tube is more stable to NEW instabilities than an underdense flux tube. The range of $k_zR$ corresponding to NEW instabilities is the same in both $V_\mathrm{ei} > 1$ and $V_\mathrm{ei} < 1$ cases (c).

Eq.~(\ref{eq:36}) shows that $\gamma_\pm$ is proportional to $\tilde{\nu}_e$. Using the estimating expression for the viscosity $\eta \approx 10^{-17}T^{5/2}$~kg\,m$^{-1}$s$^{-1}$ \citep{1986ApJ...306..730H}, we have $\tilde{\nu}_e \approx 10^{-17}T^{5/2}/\rho_ev_\mathrm{Ai}R$ in MKS units. For the typical parameters of a coronal active region, $\rho_\mathrm{e}=0.5\times 10^{-12}$~kg\,m$^{-3}$, $T=2.5\times10^{6}$~K,$v_\mathrm{Ai}=6\times10^5$~m\,s$^{-1}$, and $R=10^6$~m, we obtain $\tilde{\nu}_\mathrm{e} \approx 0.33$. Thus, the appearance and growth rate of NEW depend on the background plasma temperature and shear flow speed.
{The Reynolds number can be written as $R_e=({\omega}/{k_z})/{v_\mathrm{Ai}\tilde{\nu}_e}$, from which the valid condition for our approach is induced as $\omega/k_z\gg \tilde{\nu}_ev_\mathrm{Ai}$.}
{As discussed in Sec.~\ref{sec:2.2}, it is also necessary to consider the lower limit of $k_zR$ for the valid range of NEW instability for a given $\nu_e$: $k_zR\gg(1/R_e)$.}



\section{Conclusions and discussions}
\label{sec:conclusion}
We studied conditions for the existence of NEW surface MHD waves in a cylindrical flux tube with a shear flow in the incompressible limit. The equilibrium plasma density and shear flow experience a sharp change at the boundary of the tube. By matching the boundary condition at the tube boundary, we derived analytically dispersion relations for Alfv\'enic perturbations, and analysed dependences of the phase speed and growth rate on the plasma parameters.
Instabilities associated NEW can be excited when the shear flow speed is between the critical speed for the appearance of NEW, $U_\mathrm{c}$ and the KHI threshold, $U_\mathrm{KH}$. In other words, the steady flow shear which leads to the occurrence of NEW instabilities could be significantly lower than the KHI threshold. For example, in the long wavelength regime, {the} NEW instability threshold requires the flow speed shear several times lower than the KHI threshold. A similar result was obtained for the kink mode by~\cite{1988JETP...67.1594R} in the thin flux tube approximation. Moreover, the critical value of the flow shear for the onset of NEW instabilities should be comparable, only 20\%--40\% higher than the Alfv\'en speed inside the plasma cylinder. Such flow shears could be reached in various solar coronal plasma jets \citep[see, e.g.][]{2016SSRv..201....1R}, making them subject to NEW instabilities. Thus, NEW effects could be responsible for the occurrence of kink oscillations on a hot plasma jet, analysed by \citet{2009A&A...498L..29V}.

For the shear flow speeds lower than the KHI threshold, NEW are found to appear for all axial wave numbers.
More rigorously, the shear flow range that corresponds to NEW phenomena, $U_\mathrm{c} < U_0 < U_\mathrm{KH}$, is found to depend on the Alfv\'{e}n speed contrast inside and outside the flux tube, and also on the axial wave number of the perturbation. For all considered combinations of the parameters, the threshold value of the shear flow for the appearance of sausage NEW is lowest for the longest axial wavelengths, while the minimum value of the shear flow for kink waves is reached for the axial wavelength comparable to the diameter of the cylinder. This allows for the excitation of quasi-monochromatic perturbations by a NEW instability at a coronal jet, which is consistent with the findings of \citet{2009A&A...498L..29V}.  It is easier to excite MHD NEW in an underdense flux tube than in an overdense one, which may be used in the interpretation of kink waves obsered in supra-arcade flows \citep{2005A&A...430L..65V, 2009MNRAS.400L..85C}. It may also play an important role in the MHD wave generation and propagation in other plasma non-uniformities with field-aligned shear flows in the solar atmosphere, in particular, in the photosphere and chromosphere.

As an example of a NEW instability, we demonstrated the occurrence of dissipative NEW instability caused by finite viscosity, and found that the instability increment depends strongly on the plasma temperature and the shear flow speed. The excitation of non-axisymmetric NEW, i.e., with $m>0$, such as  kink waves, is most effective in the long wavelength limit.
Our results indicate that the omnipresence of inhomogeneous flows in the Sun's atmosphere~\citep[e.g.,][]{2018ApJ...853..145M} may lead to the effective excitation of guided MHD waves by NEW instabilities. In particular, the NEW effect may be responsible for decayless (undamped) kink oscillations~\citep[see, e.g.,][]{2016A&A...591L...5N}, which would require a dedicated study in the compressible regime typical for the solar corona. In addition, the developed model may have applications to the solar wind, the Earth's magnetotail and other plasma environments with shear flows.

{Our results are based on the assumption that the viscosity, which is assumed small, only affects the temporal behavior of the wave displacement, which may be valid in the early stage of the NEW instability. Our theory can be {tested} and its valid range can be investigated rigorously in the numerical simulations.}

\acknowledgments
The authors are grateful to the anonymous referee for his/her crucial and invaluable comments which led to a significant improvement of the manuscript.
D.J.Y. and V.M.N. acknowledge the support by the BK21 plus program through the National Research Foundation (NRF) funded by the Ministry of Education of Korea. V.M.N. acknowledges the Russian Foundation for Basic Research Grant No. 18-29-21016.

\vspace{5mm}

\newpage

\appendix

\section{Wave equation inside the flux tube}\label{sec:append_a}
Inside the flux tube, using Eqs.~(\ref{eq:1})-(\ref{eq:6}),  and assuming an axial constant magnetic field $\textbf{B}_0=(0,0,B_{0})$ and a background steady flow $U_0$ along the field,  the perturbed quantities of the radial and azimuthal velocities, magnetic field components, and total pressure, $v_{r}$, $v_{\phi}$, $b_{r}$, $b_{\phi}$, $b_{z}$, and $P(=B_0b_z/\mu_0)$, can be written as
\begin{eqnarray}
\rho_i\mathcal{F}_{u}v_{ri}&=&-P_i'+\frac{\mathcal{F}_B}{\mu_0}b_{ri},\label{eq:a1}\\
\rho_i\mathcal{F}_{u}v_{\phi{i}}&=&\frac{\mathcal{F}_B}{\mu_0}b_{\phi{i}}-\frac{1}{r}\frac{\partial P_i}{\partial\phi},\label{eq:a2}\\
\mathcal{F}_{u}b_{ri}&=&\mathcal{F}_Bv_{r{i}},\label{eq:a3}\\
\mathcal{F}_{u}b_{\phi{i}}&=&\mathcal{F}_Bv_{\phi{i}},\label{eq:a4}\\
\dot{b}_{zi}&=&-\frac{B_0(rv_{ri})'}{r}
-\frac{B_0}{r}\frac{\partial v_{\phi i}}{\partial\phi}\nonumber\\
&&+\frac{(rU_0b_{ri})'}{r}+\frac{U_0}{r}\frac{\partial b_{\phi i}}{\partial\phi},\label{eq:a5}
\end{eqnarray}
where prime and dot denote the derivative with respect to $r$ and time, respectively, and
\begin{eqnarray}
{\mathcal{F}}_{u}&=&\frac{\partial}{\partial t}+U_{0}\frac{\partial}{\partial z},~~~\mathcal{F}_B=B_{0}\frac{\partial}{\partial z}.\label{eq:a6}
\end{eqnarray}

From Eq.~(\ref{eq:a5}), we obtain the equation for $P$
\begin{eqnarray}
{\mathcal{F}}_{u}\dot{P}_i&=&-\frac{B_0(rB_0)'}{\mu_0r}\mathcal{F}_{u}v_{ri}
-\frac{B_0^2}{\mu_0}\mathcal{F}_{u}v_{ri}'
-\frac{B_0^2}{\mu_0r}\frac{\partial\dot{v}_{\phi{i}}}{\partial\phi}\nonumber\\
&&+\frac{B_0}{\mu_0r}\mathcal{F}_{u}(rU_0b_{ri})'.\label{eq:a7}
\end{eqnarray}

With Eqs.~(\ref{eq:a2})-(\ref{eq:a3}) and (\ref{eq:a7}), by applying $\mathcal{F}_u$ to Eqs.~(\ref{eq:a1}) and (\ref{eq:a7}),
we derive the wave equations for $P$ and $v_r$,
\begin{eqnarray}
\mathcal{L}_zv_{ri}&=&
-{\mathcal{F}}_{u}{P}_i',\label{eq:a8}\\
\bigg(\mathcal{L}_z-\frac{\rho_iv_\mathrm{Ai}^2}{r^2}
\frac{\partial^2}{\partial\phi^2}\bigg)\dot{P}_i&=&\\
\mathcal{L}_z\bigg[-\frac{B_{0}^2}{\mu_0r}v_{ri}
&-&\frac{B_{0}^2}{\mu_0}v_{ri}'
+\frac{B_{0}}{\mu_0r}(rU_0b_{ri})'\bigg],\label{eq:a9}\nonumber
\end{eqnarray}
where
\begin{eqnarray}
\mathcal{L}_z&=&\rho_i{\mathcal{F}}_{u}^2-\frac{B_{0}^2}{\mu_0}\frac{\partial^2}{\partial z^2}.\label{eq:a10}
\end{eqnarray}

\section{Wave equation outside the flux tube}\label{sec:append_b}
With the same axial magnetic field $B_{0}$ as inside the flux tube, assuming shear viscosity and no background flow,  and using Eqs.~(\ref{eq:1})-(\ref{eq:6}), we obtain for $v_r$, $v_\phi$, $b_r$, and $P$,

\begin{eqnarray}
\rho_eF_av_{re}&=&-\frac{2\rho_e\nu_e}{r^2}\frac{\partial v_{\phi e}}{\partial\phi}-P_e'+\frac{\mathcal{F}_B}{\mu_0}b_{re},\label{eq:b1}\\
\rho_eF_av_{\phi{e}}&=&\frac{2\rho_e\nu_e}{r^2}\frac{\partial v_{re}}{\partial\phi}
-\frac{1}{r}\frac{\partial P_e}{\partial\phi}+\frac{\mathcal{F}_B}{\mu_0}b_{\phi e},\label{eq:b2}\\
\dot{b}_{re}&=&\mathcal{F}_Bv_{re},\label{eq:b3}\\
\dot{b}_{\phi e}&=&\mathcal{F}_Bv_{\phi e},\label{eq:b4}\\
\dot{b}_{ze}&=&-\frac{B_0(rv_{re})'}{r}
-\frac{B_0}{r}\frac{\partial v_{\phi e}}{\partial\phi},\label{eq:b5}
\end{eqnarray}
where $\nu_e=\mu/\rho_e$ and
\begin{eqnarray}
{{F}}_a=\frac{\partial}{\partial t}-\nu_e\bigg(\mathcal{D}-\frac{1}{r^2} \bigg).\label{eq:b6}
\end{eqnarray}
From Eq.~(\ref{eq:b5}) we obtain the equation for $P$
\begin{eqnarray}
\dot{P}_e=K_av_{re}+K_bv_{re}'+K_cv_{\phi e},\label{eq:b7}
\end{eqnarray}
where
\begin{eqnarray}
{K}_a&=&-\frac{B_{0}^2}{\mu_0r}
,~~{K}_b=-\frac{B_{0}^2}{\mu_0},~~
{K}_c=-\frac{B_{0}^2}{\mu_0r}\frac{\partial}{\partial\phi}.\label{eq:b8}
\end{eqnarray}
Taking time derivative of Eqs.~(\ref{eq:b1})-(\ref{eq:b2}) results in
\begin{eqnarray}
\rho_eF_a\dot{v}_{re}&=&J_av_{re}+J_cv_{\phi e}-\dot{P}_e',\label{eq:b9}~~~~~\\
L_zv_{\phi e}&=&{L}_{a1}v_r+{L}_{d1}\dot{P}_e,\label{eq:b10}
\end{eqnarray}
where
\begin{eqnarray}
{J}_a&=&\frac{B_{0}^2}{\mu_0}\frac{\partial^2}{\partial z^2}
,~~{J}_c=-\frac{2\rho_e\nu_e}{r^2}\frac{\partial^2}{\partial\phi\partial t},\nonumber\\
{L}_z&=&\rho_e{{F}}_a\frac{\partial}{\partial t}-\frac{B_{0}^2}{\mu_0}\frac{\partial^2}{\partial z^2},\nonumber\\
{L}_{a1}&=&\frac{2\rho_e\nu_e}{r^2}\frac{\partial^2}{\partial\phi\partial t},~{L}_{d1}=-\frac{1}{r}\frac{\partial }{\partial\phi}.\label{eq:b11}
\end{eqnarray}
With Eq.~(\ref{eq:b10}), by applying $L_z$ to Eqs.~(\ref{eq:b7})~and~(\ref{eq:b9}), we derive the equations for $v_r$ and $P$
\begin{eqnarray}
\rho_e{L}_z{{F}}_a\dot{v}_{re}&=&
{L}_z{J}_a{v}_{re}
+\bigg({J}_c{L}_z+\frac{6\rho_e^2\nu^2}{r^4}\frac{\partial^3}{\partial\phi\partial t^2}\bigg){v}_{\phi e}~~\nonumber\\&&
-{L}_z\dot{P}_e',\label{eq:b12}\\
({L}_z-{K}_c{L}_{d1})\dot{P}_e&=&({L}_z{K}_a+
{K}_c{L}_{a1})v_{re}+{K}_b{L}_z{v}_{re}'.
\label{eq:b13}
\end{eqnarray}


\end{document}